\documentclass[12pt]{article}
\usepackage{amssymb}
\usepackage{amsthm}
\usepackage{amsmath}
\usepackage{latexsym}
\usepackage{epsfig}
\usepackage{graphicx}
\usepackage{wasysym}
\usepackage{threeparttable}
\usepackage[authoryear,comma,sort&compress]{natbib}
\usepackage{color}
\usepackage{epstopdf}
\usepackage{caption}
\usepackage{subcaption}
\newcommand{\mbv}[1]{\mbox{\boldmath$#1$\unboldmath}}
\newcommand{\mbf}[1]{\mathbf{#1}}

\newtheorem{Propriety}{Theorem}

\usepackage{hyperref}

\newcommand{\Appendix}
{
\def\thesection{Appendix~\Alph{section}}
\def\thesubsection{A.\arabic{subsection}}
}

\def\log{\hbox{log}}

\def\boxit#1{\vbox{\hrule\hbox{\vrule\kern6pt
          \vbox{\kern6pt#1\kern6pt}\kern6pt\vrule}\hrule}}

\def\bse{\begin{eqnarray*}}
\def\ese{\end{eqnarray*}}
\def\be{\begin{eqnarray}}
\def\ee{\end{eqnarray}}
\def\bq{\begin{equation}}
\def\eq{\end{equation}}
\def\bse{\begin{eqnarray*}}
\def\ese{\end{eqnarray*}}

\begin{document}
\thispagestyle{empty} \baselineskip=28pt

\begin{center}
{\LARGE{\bf Bayesian Semiparametric Hierarchical Empirical Likelihood Spatial Models}}
\end{center}

\baselineskip=12pt

\vskip 2mm
\begin{center}
Aaron T. Porter\footnote{(\baselineskip=10pt to whom correspondence should be addressed) Department of Statistics, University of Missouri-Columbia, 146 Middlebush Hall, Columbia, MO 65211, porterat@missouri.edu},
Scott H. Holan\footnote{\baselineskip=10pt  Department of Statistics, University of Missouri-Columbia, 146 Middlebush Hall, Columbia, MO 65211-6100},
Christopher K. Wikle$^2$
\end{center}
%
%
%
%
\vskip 4mm

\begin{center}
\large{{\bf Abstract}}
\end{center}
We introduce a general hierarchical Bayesian framework that incorporates a flexible nonparametric data model specification through the use of empirical likelihood methodology, which we term semiparametric hierarchical empirical likelihood (SHEL) models. Although general dependence structures can be readily accommodated, we focus on spatial modeling, a relatively underdeveloped area in the empirical likelihood literature. Importantly, the models we develop naturally accommodate spatial association on irregular lattices and irregularly spaced point-referenced data. We illustrate our proposed framework by means of a simulation study and through three real data examples. First, we develop a spatial Fay-Herriot model in the SHEL framework and apply it to the problem of small area estimation in the American Community Survey.  Next, we illustrate the SHEL model in the context of areal data (on an irregular lattice) through the North Carolina sudden infant death syndrome (SIDS) dataset.  Finally, we analyze a point-referenced dataset from the North American Breeding Bird survey that considers dove counts for the state of Missouri.  In all cases, we demonstrate superior performance of our model, in terms of mean squared prediction error, over standard parametric analyses.

\baselineskip=12pt

%
%
%

\baselineskip=12pt
\par\vfill\noindent
\par\medskip\noindent
\clearpage\pagebreak\newpage \pagenumbering{arabic}
\baselineskip=24pt
\section{Introduction}\label{sec:Intro}

The empirical likelihood (EL) dates back to the seminal work of \cite{owen1988empirical} and has become increasingly popular in recent years, as a result of \cite{owen2001}, which placed many of the fundamental concepts in a single text. Early work by \cite{qin1994empirical} greatly expanded the use of EL by placing it in the context of estimating equations. \cite{kolaczyk1994} derived general conditions for the use of estimating equations for the EL that are applicable to many types of linear, nonlinear, and semiparametric models. Many EL-type estimators have since been derived, known as Generalized Empirical Likelihood (GEL) estimators.  \cite{newey2004higher} provides an excellent overview of these estimators and their higher order properties.

\cite{lazar2003bayesian} provides evidence, by means of a simulation study, that the EL framework is appropriate for Bayesian inference. Making use of a result from \cite{monahan1992proper} that yields conditions by which a likelihood can be determined suitable for Bayesian inference, this paper initiated Bayesian research on EL and GEL estimators. \cite{schennach2005bayesian} derived a Bayesian GEL estimator by means of nonparametric priors and further extended their approach in \cite{schennach2007point}. \cite{fang2006empirical} derived the asymptotic frequentist coverage properties of the Bayesian credible intervals for the mean parameters of a wide class of EL-type likelihoods, and demonstrated undercoverage for credible intervals for parametric means generated by GEL estimators. Additional work comparing the properties of credible intervals for specific types of EL-type likelihoods can be found in \cite{chang2008bayesian}. In particular, this work demonstrates favorable coverage rates for the traditional EL of \cite{owen1988empirical}.

Bayesian hierarchical modeling (BHM) has become an expansive field. When modeling complex stochastic phenomena within the BHM framework, typically  at least three levels of model hierarchy are considered, which are the data model, process model, and parameter model \citep{berliner1996hierarchical, wikle2003hierarchical}. Subsequently, modeling typically proceeds by selecting parametric distributions for each stage of the hierarchy. As demonstrated in \cite{wiklecressie}, this framework advantageously also allows for scientifically motivated process models to be utilized at the latent stage. One aspect of this approach is that model implementation typically requires selection of an appropriate data distribution (likelihood) for the observations.

Our approach extends the general applicability of BHMs by broadly placing them in the context of the empirical likelihood. The model we propose can be viewed as a semiparametric empirical likelihood (SHEL) model and utilizes either EL estimators or GEL estimators at the data stage of the model hierarchy. Parametric process models can then be utilized to handle the potentially complex underlying dependence structures. By placing the EL in the context of Bayesian hierarchical modeling, we alleviate the issues of modeling the dependency in the observations, which is often difficult to handle in the usual observation-driven EL framework and generally utilizes restrictive blocking arguments. Specifically, we expand the BHM framework to allow empirical data models, rather than requiring the user to select a parametric structure for the data.

Hierarchical approaches to empirical likelihood have been recently considered, but still remain largely underdeveloped, with no general framework to date. \cite{chaudhuri2011empirical} proposed using the EL in a semiparametric hierarchical nested error regression model for small area estimations (SAE). The model they developed extends the traditional Fay-Herriot (FH) model \citep{fay1979estimates} to the EL framework.  Although \cite{chaudhuri2011empirical} demonstrate good model performance, their implementation utilized informative priors for some of the model parameters, and they noted sensitivity to these specifications.  The general approach they propose allows for both semiparametric and nonparametric specifications of the model for the superpopulation mean, with the nonparametric specification relying on a Bayesian nonparametric formulation (i.e., a Dirichlet process mixture with Gaussian base measure).  We pursue a more complete development of EL in the context of BHMs. The model we propose here is of independent interest and readily allows for various other hierarchical and/or dependence structures, such as temporal and/or spatio-temporal dependencies. However, for the sake of brevity, subsequent exposition focuses on spatially correlated data.

Based on blocking arguments originally developed for time series by \cite{kitamura1997empirical}, \cite{nordman2008point} developed a point referenced spatial model in the frequentist EL framework that considers variogram fitting for data collected on a regular grid, and assumes stationarity. Utilizing a similar blocking argument, \cite{nordman2008empirical} considered a observation-driven model for spatial data on a regular lattice using the EL framework that does not require stationarity.  To the best of our knowledge, hierarchical models for spatial data on an irregular lattice that explicitly account for the underlying spatial structure in the data do not exist in the current literature. A recent advancement in the spatial EL literature is \cite{bandyopadhyay2012frequency}, in which irregularly spaced spatial data is modeled using frequency domain techniques. Their framework greatly expands EL methodology for point referenced spatial data but is based on different assumptions than those presented herein and does not immediately extend to the lattice case, where distances are not uniquely defined.

The structure of this paper is as follows.  Section~\ref{sec:SSHEL} develops methodology that will be needed for the general specification of the SHEL model. Section~\ref{sec:BMF} discusses technical details related to the Bayesian estimation of the SHEL model. Two simulation studies are provided in Section~\ref{sec:SS}, whereas Section~\ref{sec:IE} presents three case studies: the FH model for SAE in the context of the American  Community Survey (ACS), the North Carolina SIDS data (areal data), and a point referenced dataset from the North American Breeding Bird survey that considers dove counts for the state of Missouri. Section~\ref{sec:Disc} provides concluding discussion.

\section{Spatial SHEL Models}\label{sec:SSHEL}

\subsection{The SHEL Framework}\label{sec:SHELFramework}

Let $\mbf{Z}$ be an $n_Z$-dimensional vector of observations, $\mbf{Y}$ be an $n_Y$-dimensional vector corresponding to an unobserved process, and $\mbv{\xi}$ be a set of parameters related to both the data model and process model. Here, $\mbf{Z}$ and $\mbf{Y}$ do not need to be of the same dimension.  For example, the observations could be mapped to the unobserved process through a matrix that accounts for change-of-support or aggregation \citep{wikleberliner2005}. However, for ease of notation, we assume $n_Z=n_Y\equiv n$, unless specified otherwise.  Further, let $[\mbf{Z}|\mbf{Y}]$ denote the conditional distribution of $\mbf{Z}$ given $\mbf{Y}$ and $[\mbf{Y}]$ denote the marginal distribution of $\mbf{Y}$. We propose a general set up for the SHEL framework that considers a data model $[\mathbf{Z}|\mbf{Y},\boldsymbol{\xi}_{D}]$, process model $[\mbf{Y}|\boldsymbol{\xi}_{P}]$, and parameter model $[\boldsymbol{\xi}]=[\boldsymbol{\xi}_{D},\boldsymbol{\xi}_{P}]$, with $[\boldsymbol{\xi}_{D}]$ being the joint prior distribution of the data model parameters and $[\boldsymbol{\xi}_{P}]$ being the joint prior distribution of the process model parameters.  The framework we propose here is not unique to spatial data, and any process model in which  $[\mbf{Y},\mbv{\xi}]$ is proper can be utilized.

The hierarchical framework that we propose is motivated by the parametric counterpart \citep[e.g.,][]{berliner1996hierarchical, wikle2003hierarchical}, but with increased flexibility from relaxing the parametric data model assumption. The SHEL structure hierarchy can be written as
\begin{eqnarray}
\nonumber \hbox{Empirical Data Model:}\,\,\, [\mathbf{Z}|\mbf{Y},\boldsymbol{\xi_D}]\\
\nonumber \hbox{Process Model:}\,\,\, [\mbf{Y}|\boldsymbol{\xi_P}] \\
\nonumber \hbox{Parameter Model:}\,\,\, [\boldsymbol{\xi_D},\boldsymbol{\xi_P}],
\end{eqnarray}
where the underlying distribution $[\mathbf{Z}|\mbf{Y},\boldsymbol{\xi}_D]$ is assumed to have two finite moments. Critically, we further assume $E(\mbf{Z|Y,\mbv{\xi}_D})=g(\mbf{X}\mbv{\beta}+\mbf{Y})$ and $E(\mbf{Z}^2|\mbf{Y},\mbv{\xi}_D)=h(\mbf{X}\mbv{\beta}+\mbf{Y})$ for $g$ and $h$ known, with $\mbf{X}$ being an $n \times m$ design matrix of fixed and known covariate information. These relationships will serve to inform a set of estimating equations utilized in estimating the parameters of the empirical data model.

When utilizing the SHEL framework, $[\mathbf{Z}|\mbf{Y},\boldsymbol{\xi}_D]$ will be modeled empirically, using the EL. As a result, our approach typically allows for the data to be modeled directly. This avoids the need to identify an appropriate transformation in order to model data that do not follow a known distribution and allows for model development to proceed in cases where no appropriate transformation exists. The spatial SHEL model we propose creates a unifying model for empirical likelihood-based Bayesian hierarchical spatial modeling.

One of the main advantages of working in the hierarchical paradigm with an EL data model is the ability to introduce conditional independence in a natural way, specifying the dependence structure at a higher level in the model hierarchy. That is, dependence among outcomes in a spatial (and/or temporal) setting is handled by conditioning on a latent spatial (and/or temporal) process.  By taking a conditional approach, the original formulation of the EL, which assumes independent and identically distributed (i.i.d.) observations, becomes immediately applicable -- although the assumption of independent observations could be relaxed \citep[e.g., see][Chapter 4]{owen2001}.  In other words, the SHEL framework effectively utilizes the conditional model specification inherent to BHMs to extend the applicability of the EL to a broad range of analyses. In doing so, we alleviate some of the strict assumptions often required of the blocking arguments used in EL modeling of dependent data, such as those in \cite{kitamura1997empirical}, \cite{nordman2008point},  \cite{nordman2008empirical}, and \cite{kaiser2012blockwise}.

\subsection{Empirical Likelihood}

The use of estimating equations in the EL framework \citep{qin1994empirical} has recently been used in the FH model by \cite{chaudhuri2011empirical} and represents an attractive way to employ EL in the BHM framework. Generally, the EL of a vector of functionals $\mbv{\theta}=\{\theta_1,\ldots,\theta_R\}$ given independent and identically distributed observations $Z_1,\ldots,Z_n$, can be computed as
\begin{equation}
L(\mbv{\theta})\propto \prod_{i=1}^{n} {w}_{i}(\mbv{\theta})
\label{eq:EL}
\end{equation}
where $L(\mbv{\theta})$ is maximized over the simplex
\begin{equation}
 W_{\theta}=\left\{ \sum_{i=1}^{n} w_{i} =1; w_{i}>0 \text{ for all $i$ }; \sum_{i=1}^{n}w_{i}m_{j}(z_i,\theta_i)=0 \hbox{ for all $j$} \right\}
 \label{eq:EESimplex}
\end{equation}
and $R$ is the number of functionals to be estimated.  Here, for $i$ in $1,\ldots,n$, $\{m_{j}(z_i, \theta_i)\}_{j=1,...,J}$ are a set of $J$ estimating equations and $\boldsymbol{\theta} \in \mathbb{R}^{J}$ are of the form $k_j(\sum_{i=1}^{n}w_{i} z_{i})=\theta_{j}$, for known functions $k_j(\cdot)$, where we have assumed $J=R$, i.e., that unstructured $\boldsymbol{\theta}$ is not under- or overspecified. Without covariate information, one cannot estimate more parameters than the number of estimating equations. However, \cite{chaudhuri2011empirical} suggest utilizing structured $\mbv{\theta}$, by which each location $i=1,\ldots,n$ has a unique mean and variance. Covariate information is then used to provide structure to a set of mean parameters $\{\theta_1,\ldots,\theta_n\}$, where $\theta_i$ is modeled based on auxiliary information $\mbf{x}_i$. This covariate information allows the dimension of $\mbv{\theta}$ to be greater than $J$. The estimating equations \cite{chaudhuri2011empirical} suggest have the form
\begin{eqnarray}
\nonumber & &\sum_{i=1}^{n} w_{i}\{z_{i}-\theta_{i}\}=0 \\
\label{eq:EE}
& &\sum_{i=1}^{n} \{w_{i}(z_{i}-\theta_{i})^{2}/V(\theta_{i})\}-1 =0,
\end{eqnarray}
which are derived based on the exponential family. In the exponential family we define $\theta_{i}$ to be mean of $Z_{i}$ and $V(\theta_{i})$ to be the variance of $Z_{i}|\theta_{i}$. These easily extend to the GEL framework, but $V(\theta_{i})$ is no longer properly considered a variance, instead serving as a scale parameter.

In the SHEL framework, $\theta_i$ will denote the conditional mean of $Z_i|Y_i$. The estimating equations approach is natural for the SHEL framework because one can compute the EL based on known formulas given proposed values for $\{\theta_i$\}.  When utilizing the estimating equation approach to the EL, the model weights can be computed as
\begin{eqnarray}
w_{i}=\frac{1}{n}\left(\frac{1}{1+\sum_{j=1}^J\lambda_{j}m_{j}(z_{i},\theta_i)}\right),
 \label{eq:ELEEweights}
\end{eqnarray}
where $\lambda_j, j=1,\ldots,J$ satisfies
\begin{eqnarray}
\nonumber \sum_{i=1}^{n} \frac{m_{j}(z_{i},\theta_{i})}{1+\sum_{j=1}^J\{\lambda_j m_j(z_{i},\theta_i)\}}=0
\end{eqnarray}
for all $j$, and $\{z_i\}$ denote the observations. Clearly, these weights are monotone in each element of $\boldsymbol{\lambda}=\{\lambda_1,\ldots,\lambda_J\}$.

The likelihood can be extended to a set of GEL estimators \citep{smith1997alternative} by the function
\begin{equation}
L(\theta) \propto \prod_{i=1}^{n} \widehat{w}_{i},
\label{eq:GEL}
\end{equation}
where $\widehat{w}(\mbv{\theta})=$ argmax$_{W_{\theta}}\sum_{i=1}^{n} f\{w_{i}(\mbv{\theta})\}$ for a known function $f_\theta(w_i)$. Two notable choices include  $f_\theta(w_i)=\log(w_{i})$, which is the traditional EL function first introduced by \cite{owen1988empirical}, and $f_\theta(w_i)=-w_{i} \log(w_{i})$, which was introduced by \cite{schennach2005bayesian} and represents the exponentially tilted empirical likelihood (ETEL) estimator. Henceforth, we utilize only the traditional EL of \cite{owen1988empirical} throughout the methodological development, but note that other choices of $f_\theta(\cdot)$ in the GEL family could also be used.

An important observation of \cite{chaudhuri2011empirical} is that the non-analytic form of the posterior distributions introduced by the EL makes verification of propriety of these models difficult. Therefore, improper priors should generally not be used in the SHEL framework, as only proper priors can guarantee propriety of the posterior parameter distributions.

\subsection{Lattice Priors for the SHEL Framework}

Intrinsic Gaussian Markov Random Fields (IGMRFs) \citep{rue2005gaussian}, such as the intrinsic conditional autoregressive model (ICAR) \citep{besag1991}, may seem to be a poor choice for a SHEL prior due to the impropriety implicit to these models. However, recent developments in lattice priors allow for modification of the ICAR to yield a proper prior, while avoiding some of the common difficulties of proper CAR models.

A common ICAR model specification is given by
\begin{equation}
\nonumber Y_{i} \sim N\left(\sum_{j \in ne(i)} \left\{\frac{ b_{ij}}{ \sum_{j \in ne(i)} b_{ij}} y_{j} \right\}, \frac{\sigma^2}{ \sum_{j \in ne(i)} b_{ij}}\right),
\end{equation}
where $b_{ij}=1$ if locations $i$ and $j$ are neighbors and $0$ otherwise, and $j \in ne(i)$ indicates that locations $i$ and $j$ are neighbors.  This yields a probability density function for $\mbf{Y}=(Y_1,\ldots,Y_n)'$ given by
\begin{eqnarray}
\nonumber \pi(\mbf{Y}=\mbf{y}) \propto \exp \left\{-\frac{1}{2} \mbf{y}' \tau (\mbf{B}_{+}-\mbf{B}) \mbf{y}\right\},
\end{eqnarray}
where $\mbf{B}$ is a matrix with $\{B\}_{(i,j)}=b_{ij}$ and $\mbf{B}_{+}$ is a diagonal matrix with $\{B_+\}_{i,i}=\sum_{j \in ne(i)} b_{ij}$. The IGMRF specification of this model would therefore imply the log-density of $\mbf{y}$ as
\begin{equation}
\pi(\mbf{Y}=\mbf{y})=-\frac{n-1}{2} \log(2 \pi) +\frac{1}{2} \sum_{i=1}^{n-1} \log(\lambda_{i})-\frac{1}{2}\mbf{y}'\tau(\mbf{B}_{+}-\mbf{B})\mbf{y},
\end{equation}
where $\lambda_{1}\geq \cdots \geq \lambda_{n}$ are the ordered eigenvalues of $(\mbf{B}_{+}-\mbf{B})$.

Because $(\mbf{B}_{+}-\mbf{B}) \mathbf{1}=\mathbf{0}$, where $\mathbf{1}$ is a vector of ones, we see that the precision matrix $\tau(\mbf{B}_{+}-\mbf{B})$ is singular, and the ICAR can only be utilized as an improper prior. One possible solution is to modify the matrix $\tau(\mbf{B}_{+}-\mbf{B})$, by adding a spatial dependency parameter. For this ICAR parameterization, the matrix $\tau(\mbf{B}_{+}-\rho \mbf{B})$ is guaranteed to be positive definite for $\rho \in (-1,1)$. However, there are a two major drawbacks to introducing a spatial dependency parameter $\rho$. First, $\rho$ must be quite large to generate significant spatial dependency, and a uniform prior distribution often leads to diffuse posterior distributions for $\rho$. Second, \cite{wall2004close} notes undesirable properties of the pairwise correlations of the locations on an irregular lattice as $\rho$ is varied throughout the space (-1,1).

\cite{hughes2010dimension} utilize an orthogonalization argument derived in \cite{reich2006effects} by considering orthogonal spatial smoothing using a generalized Moran basis. \cite{hughes2010dimension} smooth orthogonal to $\mbf{X}$ by considering an eigenvector basis of $\mbf{P}_{c}\mbf{B}\mbf{P}_{c}$ for the latent process space, where $\mbf{P}_c=\mbf{I}-\mbf{X}(\mbf{X}'\mbf{X})^{-1}\mbf{X}'$. This allows orthogonal smoothing to $\mbf{X}$ while accounting for the underlying lattice structure of the data. In our formulation of a SHEL model on a lattice, we utilize this structure. We define $\mbf{M}$ as an $n\times q$ matrix with the columns being the eigenvectors corresponding to the $q$ largest nonzero eigenvalues of the matrix $\mbf{P}_{c}\mbf{B}\mbf{P}_{c}$. The process $\mbf{Y}_n$ can then be modeled in a rank-reduced form, $\mbf{Y}_n$=$\mbf{M}_{n \times q} \mbf{Y}^*_q$, where $\mbf{Y}^*_q$ is the rank-reduced process. This model is useful because, under weak conditions, the prior of \cite{hughes2010dimension} yields a proper prior that respects the underlying lattice without the need to introduce new parameters. We now provide a sufficient condition for $\mbf{M}'(\mbf{B}_+-\mbf{B})\mbf{M}$ to be positive definite:
\begin{Propriety}
Consider a Bayesian hierarchical model in which the data model has two finite moments $E(\mbf{Z|Y},\mbv{\xi}_D)=g(\mbf{X}\mbv{\beta}+\mbf{M}\mbf{Y}^*)$ and $E(\mbf{Z}^2|\mbf{Y},\mbv{\xi}_D)=h(\mbf{X}\mbv{\beta}+\mbf{M}\mbf{Y}^*)$ with $g$ and $h$ being known functions. Let the process $\mbf{Y}^*$ be given a \cite{hughes2010dimension} prior of the form $\pi(\mbf{Y}^*=\mbf{y}^*) \propto \tau^{q/2} \exp\{-\frac{1}{2} \tau \mbf{y}^{*'}\mbf{M'} (\mbf{B}_{+}-\mbf{B}) \mbf{M} \mbf{y}^*\}$, where $\mbox{rank}\,(\mbf{M}) \leq n-1$. Assume that $\mbf{B}$ is the adjacency matrix for a first order IGMRF (i.e., $\mbox{rank}\,(\mbf{B})= n-1$). Then, a sufficient condition for $\mbf{M}'(\mbf{B}_+-\mbf{B})\mbf{M}$ to be positive definite is that the design matrix $\mbf{X}$ contains a column corresponding to an intercept term.
\end{Propriety}
A proof of Theorem 1 can be found in Appendix A.  Theorem 1 implies that the \cite{hughes2010dimension} lattice prior will yield a proper prior suitable for use with EL methods on a lattice whenever one includes an intercept term in the design matrix. The main advantage of this model over other lattice priors is that this basis simultaneously allows for dimension reduction. Let $\lceil x\rceil$ denote the the ceiling of $x$ -- the smallest integer greater than or equal to $x$.  The recommendation of \cite{hughes2010dimension} is that the eigenvectors associated with the largest $q=\lceil0.1n\rceil$ eigenvalues of the matrix $\mbf{P}_{c} \mbf{B} \mbf{P}_{c}$ are typically sufficient to allow accurate estimation of the fixed effects, though there is some sensitivity to the actual proportion used.
In our analyses, which are of much lower dimensionality than those considered in \cite{hughes2010dimension}, we have found that the prediction is markedly better in terms of mean squared prediction error (MSPE) when we utilize every eigenvector of $\mbf{P}_{c} \mbf{B} \mbf{P}_{c}$ associated with a positive eigenvalue. This strategy leads to substantially decreased computation time in the SHEL framework, along with simpler tuning of the Markov chain Monte Carlo (MCMC) algorithms employed in this model relative to the full-rank implementation.

\section{Bayesian Model Estimation}\label{sec:BMF}
EL computation is well established. As early as 2001, several methods had been developed \citep{owen2001}, with additional methods building off of this early research. \cite{chen2002using} is notable in that it provides a method for computing the EL with guaranteed convergence. We propose a straightforward approach that allows the built-in optimization functionality of the R programming language \citep{Rlanguage} to be utilized for fast computation.

An issue to overcome is selecting starting parameter and latent values that allow the EL to be computed. We propose setting the process model values to zero, and utilizing the maximum empirical likelihood estimates (MELEs) of the fixed effects as the starting values of the chain. The $gmm$ package in R \citep{chausse2010computing} can be used to rapidly obtain these starting values. MCMC computations can then proceed via standard Metropolis-Hastings methodology for any parameter appearing in the estimating equations for the EL portion of the model.

In the case where the model defined by the estimating equations approach to EL as outlined in (\ref{eq:EE}) is not over- or under-determined, the solution for $\boldsymbol{\lambda}=\{\lambda_1,\ldots,\lambda_J\}$, if it exists, is unique for a given value of $\boldsymbol{\theta}=\{\theta_1,\ldots,\theta_R\}$. The EL constraints for $\boldsymbol{\lambda}$ are $\left\{\sum_{i=1}^{n} w_{i} m_{j}(z_i,\theta_{j})=0\right\}_{j=1,...,J}$. This structure can be exploited by using the $optim$ function in R  in order to find the minimum of $\sum_{j=1}^{J} \{\sum_{i=1}^{n} w_{i} m_{j}(z_i,\theta_{j})\}^{2}$. If the value of this objective function is zero, we can verify that the solution for $\{\lambda_1,\ldots,\lambda_J\}$ yields a set of weights $\{w_i,i=1,\ldots,n\}$ in the simplex of (\ref{eq:EESimplex}), by checking that $\sum_{i=1}^{n} w_{i}=1$ and that $w_{i}>0$ for all $i$. When these conditions are met, we have the value of the EL as $\prod_{i=1}^{n} w_{i}$. When using the $optim$ function, which is the default fitting method for the $gmm$ package, one must decide on a numerical threshold for deciding when $\left\{\sum_{i=1}^{n} w_{i} m_{j}(z_i,\theta_{j})=0\right\}_{j=1,...,J}$ and $\sum_{i=1}^{n} w_{i}=1$ are satisfied. We have had success in evaluating this term by considering $\left\{\sum_{i=1}^{n} w_{i} m_{j}(z_i,\theta_{j})<\epsilon\right\}_{j=1,...,J}$ and $(\sum_{i=1}^{n} w_{i})-1< \epsilon$ where $\epsilon=5\times10^{-3}$.

Because the data model is non-analytic, Gibbs sampling is not possible for any of the parameters, as none of the full conditional distributions are of standard form. Therefore, we utilize Metropolis-Hastings within Gibbs (MH) sampling for all of the model parameters. Specifically, for our analyses, we use a random walk MH sampling algorithm having Gaussian proposals with variances tuned based on the empirical covariances from a pilot chain \citep{gelman2013bayesian}. An example of the algorithm can be found in Appendix B.

\section{Simulation Studies}\label{sec:SS}
Of particular interest is the performance of the SHEL paradigm in spatial prediction, and so we conduct a simulation study to assess the predictive performance of the SHEL framework as compared to parametric models.

\subsection{Study 1: The SHEL Fay-Herriot Model}\label{SimFH}
The FH model \citep{fay1979estimates} is a SAE model and can be written as
\begin{eqnarray}
 \nonumber Z_{i}=\theta_{i}+\epsilon_{i}\\
 \theta_{i}=\mathbf{x}_{i}' \mbv{\beta} + y_{i},
 \label{eq:FH}
\end{eqnarray}
where $Z_{i}$ is a design unbiased survey estimate of $\theta_{i}$, the superpopulation parameter of interest at location $i$, and $\epsilon_{i}$ is a spatially referenced sampling error with mean zero and known variance $\sigma_{i}^{2}$.  Auxiliary information at location $i$ is denoted by $\mathbf{x}_{i}$, and $\mathbf{y}=(y_1,\ldots,y_n)'$ denotes a vector of spatially referenced random effects.

Additionally, one typically assumes that, for $i=1,\ldots,n$, $\epsilon_i$ are independent and that $\mbv{\epsilon}=(\epsilon_1,\ldots,\epsilon_n)'$ follows a multivariate normal distribution. \cite{chaudhuri2011empirical} employed a Bayesian nested error regression in the FH framework that relaxed this assumption. Their analysis is demonstrated using two possible priors on $\mathbf{y}$. The first prior is an independent and identically distributed (i.i.d.) Gaussian distribution, whereas the second prior is a Dirichlet process (DP) prior with a Gaussian base measure. They note sensitivity in their analysis to the prior specification of the hyperparameters in the prior for $\mathbf{y}$, as well as to the prior for $\beta_{0}$ -- the fixed effect associated with the intercept in Equation (\ref{eq:FH}). The actual estimating equations we utilize in the EL for estimating $\{\mbv{\beta},Y\}$ are:
\begin{eqnarray*}
& &\sum_{i=1}^{n} w_{i}\{z_{i}-\theta_{i}\}=0 \\
& &\sum_{i=1}^{n} \{w_{i}(z_{i}-\theta_{i})^{2}/\sigma_i^2\}-1 =0.
\end{eqnarray*}

In the simulation study presented here, we compare the prediction of the SHEL FH model and the independence model of \cite{chaudhuri2011empirical} on data that behave similar to those of our data analysis in the FH analysis of Section~\ref{data:SHELFH}. We do not utilize their DP prior model due to concerns of computational considerations associated with repeated estimation within a full simulation study and the fact that the DP process model performs similar to the independence model in the analysis of \cite{chaudhuri2011empirical}. To simulate data, random effects $y_i$ are generated based on a \cite{hughes2010dimension} lattice prior with a precision parameter equal to the posterior mean of $\tau$ in the analysis of Section~\ref{data:SHELFH}. Then data model weights $\{w_i\}$ are generated based on the posterior means of the fixed effects parameters of that analysis. This gives an EL to generate data that will have similar properties to the data in Section~\ref{data:SHELFH}. We generate 125 datasets in this way and perform a leave-one-out MSPE analysis on each dataset. For each location within a given dataset, the model is run for 11,000 iterations, with 1,000 iteration discarded as burn-in (i.e., 10,000 used for our analysis). To assess convergence, we visually inspect a random subset of sample chains from the $125\times115$ analyses and note that no lack of convergence was detected.

For the independence prior, we used the specification $y_i \sim \hbox{N}(0,A)$ with $A\sim\hbox{IG}(1,1)$, $\beta \sim \hbox{N}(\beta^*, g^{-1}A I_2)$. The constant $g$ represents Zellner's g prior \citep{zellner1986bayesian}, here set to 10. The prior means, $\beta^*$, are the weighted least squares (WLS) estimates from a regression of the auxiliary information on the data assuming no latent effects are present. These prior specifications represent an identical formulation as in \cite{chaudhuri2011empirical}. For comparison, the SHEL model utilizes the Moran basis, with the vague priors $\tau \sim \hbox{Gamma}(1,1)$, $\beta \sim \hbox{N}(\beta^*, g^{-1}\tau^{-1} I_2)$, and $\mbf{y} = \mbf{M} \mbf{y}^*$ where $\mbf{y}^{*} \sim \hbox{N}(0, \tau \mbf{M}'\{\mbf{B}_{+}-\mbf{B}\} \mbf{M})$.

We define MSPE as $\sum_{i=1}^{44} (Z_i-\widehat{Z}_{(-i)})^2/44$ with $\widehat{Z}_{(-i)}$ being the prediction at location $i$ when the data at location $i$ is treated as missing. Over all 125 simulated datasets, the SHEL FH model provides an average MSPE of 0.163, while the independence model of Chaudhuri and Ghosh (2011) provides an average MSPE of 0.239. This represents a 31.6\% average reduction in MSPE. Notably, we see similar results in terms of MSPE reduction in Section~\ref{data:SHELFH}, and the results corroborate one another.

\subsection{Study 2: Breeding Birds}\label{SimBirds}
To illustrate an example with continuous spatial reference and non-Gaussian data we utilize the example described in \cite{gelfand2010handbook}.  In particular, the simulation study we perform is designed similar to the North American Breeding Bird Survey (``Dove") data analysis performed of Section~\ref{data:Birds} and, to assess the performance of the SHEL model relative to a parametric specification, we use the following model for comparison
\begin{eqnarray}
\nonumber & &Z(s_i)|\lambda(s_i) \sim \hbox{\textit{ind} Poisson}(\lambda(s_i)),\hbox{ }i=1,\ldots,n; \\
\label{eq:Birds}
 & &\log\{\lambda(s_i)\}=\beta+y(s_i).
\end{eqnarray}
We modeled $\mbf{y}=(y_1(s_1),\ldots,y_n(s_n))'$ as multivariate Gaussian with mean zero and covariance function $\sigma_y^2\, r(s_i,s_j;\phi)$, where $r(s_i,s_j;\phi)=\hbox{exp}(-||s_i-s_j||/\phi)$.  We placed a $N(0,100^2)$ prior on $\beta$, a $\mbox{Unif}(0.01,100)$ prior on $\sigma_{y}^{2}$ and, similar to \cite{gelfand2010handbook}, a $\mbox{Unif}(0,4)$ prior on $\phi$.

The estimating equations for the SHEL model are based on the identities $\theta_i=V(\theta_i)=\hbox{exp}\{\beta+y(s_i)\}$ in Equation (\ref{eq:EE}). This is a SHEL specification based on an overdispersed Poisson model, where we have the conditional mean, $\theta_i$, and variance of $Z_i|\theta_i$ equal.

We assess the predictive performance of the model by means of a leave-one-out mean squared prediction error MSPE experiment. In order to generate data that have similar properties to the Dove data, we first analyzed the data according to the SHEL model we propose. Next, we computed the posterior means, $\widehat{\beta}$ and $\widehat{\mbf{y}}$, from this analysis. These values were then used to compute average weights $\{w_i\}$ which correspond to an EL based on the posterior parameter means. These weights were then used, in turn, to generate new data. New random effects were generated from the spatial prior used in the analysis with $\sigma_y^2$ and $\phi$ set at their respective mean posterior values in the analysis in Section~\ref{data:Birds}. We generated 250 datasets in this way and performed a leave-one-out MSPE experiment in which we analyze each dataset 44 times, each time with a different location left out of the analysis. For each dataset, each analysis was run for 11,000 iterations, with 1,000 iterations for burn-in, resulting in 10,000 MCMC iterations which were used for analysis.  We visually inspected all 47 sample chains (44 random effects and 3 parameters) for 10 random analyses and found no evidence of non-convergence.

We define MSPE identically to that of Section~\ref{SimFH}. The SHEL model yields a MSPE of 331.4 when averaged across all 250 simulations, while the previously proposed Poisson model of \cite{gelfand2010handbook} yields a MSPE of 400.0. This constitutes a 24.7\% average MSPE reduction and strongly indicates that the SHEL model performs superior in this context.

\section{Case Studies}\label{sec:IE}
\subsection{A SHEL Fay-Herriot Model}\label{data:SHELFH}

In our FH analysis, we consider the parameter of interest to be the 2010 five year period estimate of mean per capita income in Missouri counties, obtained from the American Community Survey (ACS) (www.census.gov/ACS), which was scaled by 10,000 for numerical stability.  We utilize the percentage of unemployed individuals in each county as auxiliary information, also obtained from the ACS. The data are not normally distributed, and neither a log nor a Box-Cox transformation yielded normality.

For the SHEL analysis, the prior on $\mathbf{y}^*$ (the reduced-rank process) is taken as $N(0,  \{\tau \mathbf{M}' (B_{+}-B)\mathbf{M}\}^{-1})$, where $\mathbf{M}$ is a matrix that contains the eigenvectors of the Moran basis associated with the positive eigenvalues of the matrix $\mbf{P}_c \mbf{B} \mathbf{P}_c$ as columns. We compare our model to the independence model of \cite{chaudhuri2011empirical}, and a model using the DP prior. For these data, there was no available transformation that satisfied the normality assumption of the data, but we perform a na\"ive parametric FH analysis modeling $\mbv{\epsilon}$ as independent and normally distributed random errors for comparison.

The prior for the DP process prior was $y_i|G \sim G$, $G|A \sim \hbox{DP}(\alpha, \mathcal{G})$, where $\alpha\equiv 1$ as in \cite{chaudhuri2011empirical}  and $\mathcal{G}$ represents a Gaussian base measure. For computational reasons, we approximate the DP prior using a finite mixture of normals. For these data, we considered possible cluster counts of 20, 50, and 115 (the full data size), and found our results to be robust in terms of MSPE to the number of clusters we select $a$ $priori$. However, Chaudhuri and Ghosh (2011) utilize an informative prior on $A$, and we note substantial sensitivity to this prior specification for our data. We assumed several prior specifications for $A$ in their framework, and we present the results for $A \sim IG(2,1000)$, which yielded the lowest MSPE of any prior we tried (MSPE of 0.128). The prior specification $A\sim\hbox{IG}(2,10)$, which is of similar strength to that of \cite{chaudhuri2011empirical}, yielded an MSPE of 0.182, which was worse than the parametric analysis. Additionally, we attempted $A \sim IG(1,1)$, which is more vague than the priors used by \cite{chaudhuri2011empirical}, and found that it yielded sample chains with questionable convergence, though it yielded converged chains for the independence prior of \cite{chaudhuri2011empirical} and for the SHEL model. All other priors in these analyses were set identical to the simulation study in Section~\ref{SimFH}.  In order to assess the relative importance of the spatial structure, we additionally consider a spatial parametric FH model with the exact same prior specifications as our SHEL model but with a Gaussian data distribution.

Table~\ref{TA:FH} reports summary statistics for the posterior distributions of these models, as well as the mean posterior variances for the mean posterior predicted variances for $\mbv{\theta}=\{\theta_1,\ldots,\theta_n\}$. All model results are based on 11,000 MCMC iterations with the first 1,000 iterations discarded for burn-in (i.e., 10,000 iterations total).  Convergence was assessed through visual inspection of the sample chains, with no deviations from convergence detected.

\begin{figure}
\caption{\baselineskip=10pt The difference of the squared deviations $(Y_i-\widehat{Y}_{(-i)})^2$ for each location of estimated per capita income for (a) the SHEL model versus the Chaudhuri and Ghosh (2011) independence model, (b) the SHEL model versus the Chaudhuri and Ghosh (2011)DP model, (c) the SHEL model versus the parametric model. The square represents Kansas City, MO and the triangle represents St. Louis, MO.}
\label{Fi:FH}
\begin{center}
\begin{tabular}{cc}
\includegraphics[width=60mm, height=60mm,angle=-90]{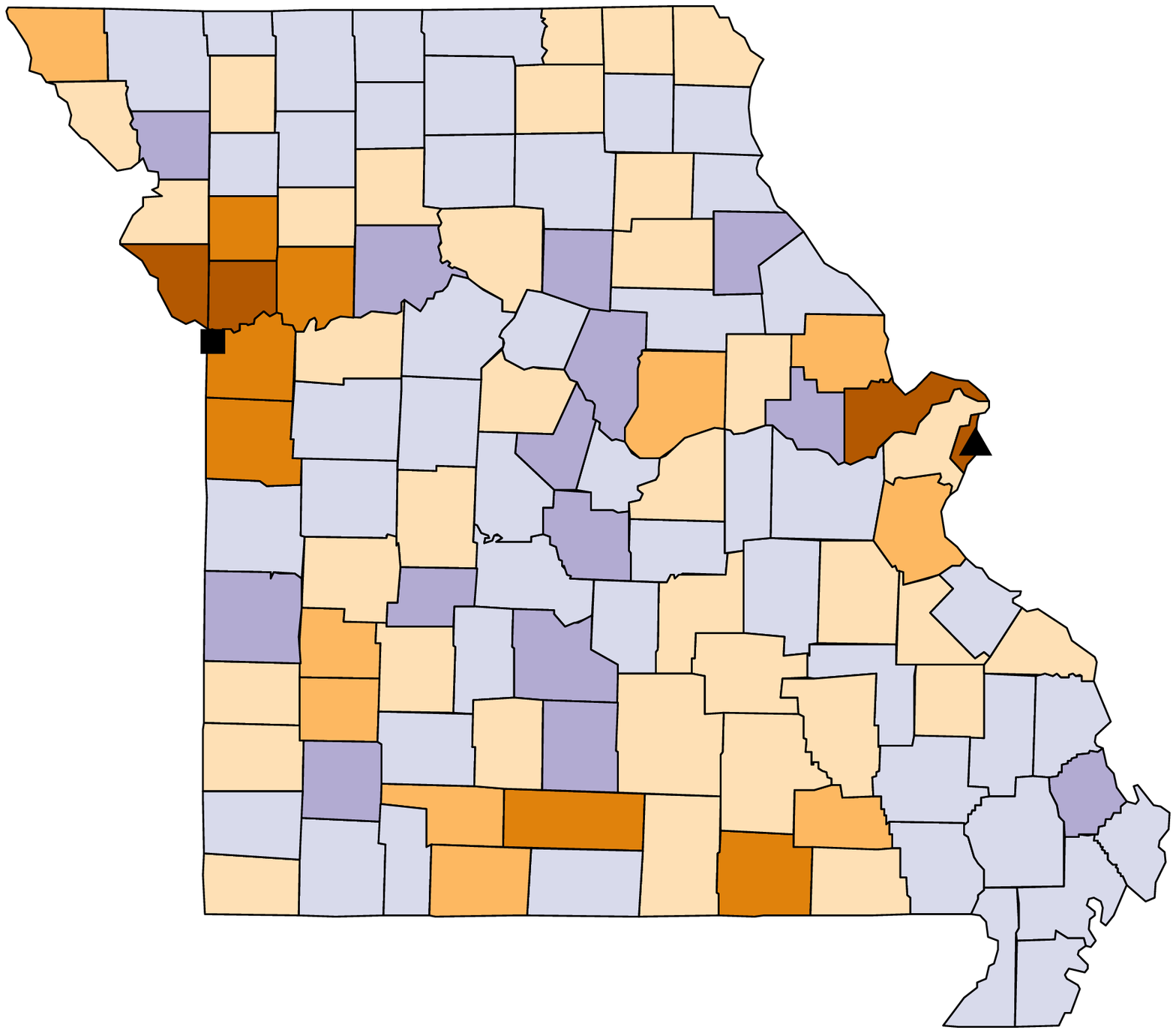} &
\includegraphics[width=60mm, height=60mm,angle=-90]{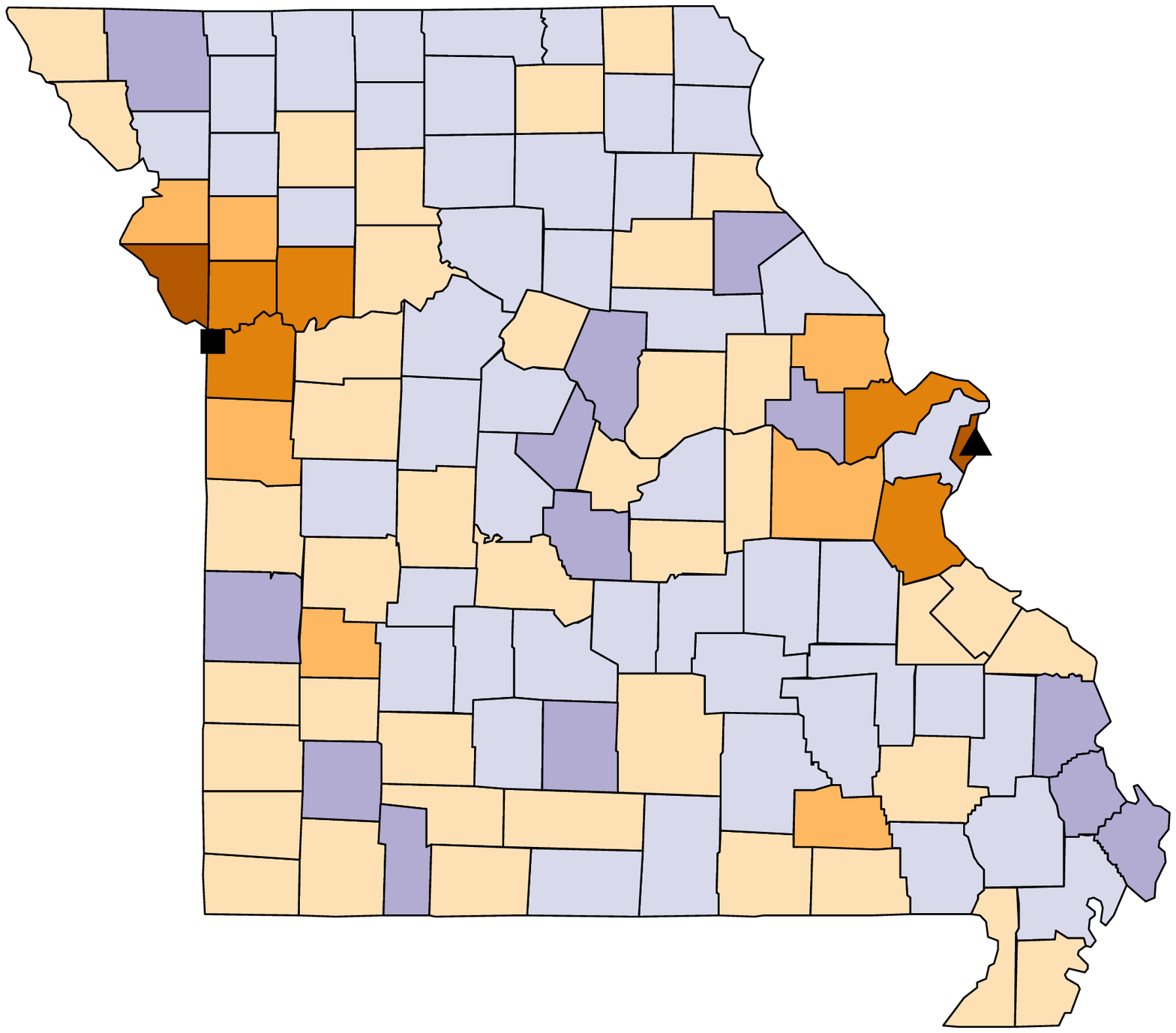} \\
 (a) & (b)\\
\includegraphics[width=60mm, height=60mm,angle=-90]{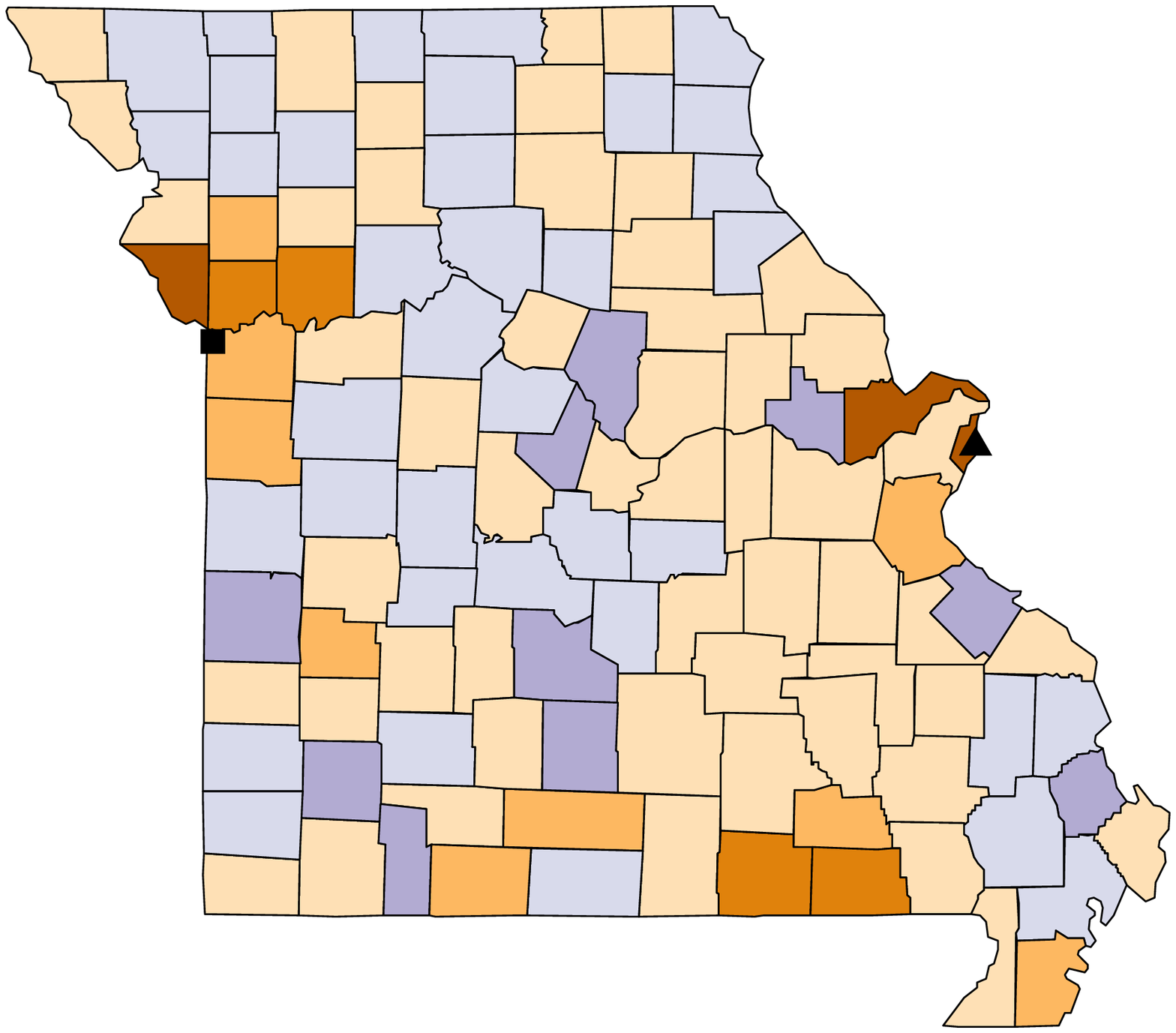} &
\includegraphics[width=60mm, height=60mm,angle=-90]{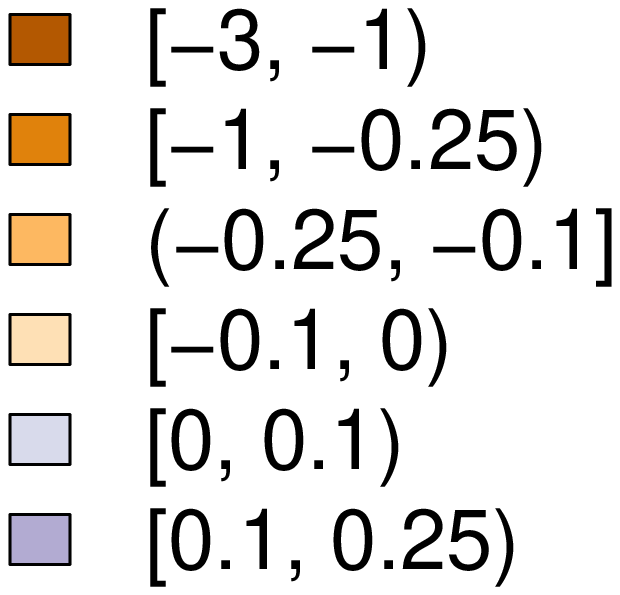}\\
 (c) & \\
\end{tabular}
\end{center}
\end{figure}

We additionally performed a leave-one-out MSPE analysis for each model. The parametric model performs nearly as well in terms of MSPE as the models of \cite{chaudhuri2011empirical}. The DP prior model of \cite{chaudhuri2011empirical}, performs nearly equivalently to their independence model in terms of MSPE.  In summary, the SHEL model, which explicitly accounts for the spatial correlation in these data, performs markedly better than all three other models, and yields a MSPE of 0.066, while the best fitting model of \cite{chaudhuri2011empirical} yields an MSPE of 0.128, which is a reduction in MSPE of 48.4\%. These results strongly indicate that the SHEL model with the Hughes and Haran (2013) lattice prior is the preferred model for these data. Additionally, the spatial parametric FH model yielded a MSPE of 0.076, underscoring the importance of accounting for the spatial correlation in these data. The differences in MSPE for each location are plotted spatially in Figure~\ref{Fi:FH} and clearly illustrate that the SHEL model provides estimates that deviate less from the observed data in the high population areas near St. Louis, MO and Kansas City, MO. These cities greatly influence the surrounding areas, and the explicit spatial autocorrelation embedded in the SHEL FH model greatly aids in the estimation of these areas. Due to the similarity in spatial performance, the spatial parametric FH is not shown in this figure.

\subsection{The North Carolina SIDS dataset} \label{data:SIDS}
The North Carolina Sudden Infant Death Syndrome (SIDS) dataset is a frequently analyzed areal dataset in spatial modeling. We utilize the data collected over the period from 1974--1978. After accounting for the counts of live births in North Carolina, there is still a significant clustering of events \cite[e.g.,][]{getis1992, kulldorf1997}. For this particular dataset, several parametric models have been considered.  For example, \cite{symons1983} first attempted to model the spatial structure in these data based on high risk and low risk populations. More recent work has considered models with more explicit formulations. The parametric model we utilize is
\begin{eqnarray}
\nonumber & &Z_i|\lambda_i \sim \hbox{Poisson}(\lambda_i),\hbox{ }i=1,\ldots,n; \\
\nonumber & &\log(\lambda_i)=log(E_i)+x_i' \mbv{\beta}+y_i,
\end{eqnarray}
which is suggested by \cite{cressie1989spatial}. Here $E_i$ is the expected SIDS count in each county, which is computed as $N_i \{\sum_{i=1}^{n}(Z_i)/\sum_{i=1}^{n}(N_i)\}$, where $N_i$ is the total number of births in county $i$. We utilize an intercept, and the proportion of births in each county that resulted in non-white children as covariates. A Tukey-Freeman transformation was applied to the proportion of non-white births, as suggested by \cite{cressie1989spatial}. This data is well modeled by an overdispersed Poisson distribution, with the exception of a single extreme outlier, which is Anson county. In the analysis performed by \cite{cressie1989spatial}, this location was left out of the analysis. More recently, \cite{sengupta2013} dealt with this outlier in the empirical Bayesian hierarchical model setting by modeling the data through a non-stationary spatial process over 13 regions, where the correlation between these spatial regions was built on Euclidian distances.  In this analysis, Anson county was considered its own region and included in the model.  Since our goal is to demonstrate the robustness of the SHEL model (in terms of MSPE) to extreme outliers, we compare our approach to \cite{cressie1989spatial}, as they removed the outlier from their analysis.  That is, we propose a relatively simple way to handle this outlier: depart from the Poisson distribution, and use EL methods to obtain estimates for $\{\mbv{\beta}, \mbf{y}\}|\mbf{Z}$. We propose estimating equations based on the identities $\theta_i=V(\theta_i)=\hbox{exp}(x_i' \mbv{\beta}+y_i)$, linking the SHEL model to the Poisson distribution.

We compare the SHEL methodology to the overdispersed Poisson suggested above, but with all of the locations considered in the analysis. For both the SHEL and parametric models, $\mbf{y}=(y_1,\ldots,y_n)^\prime$ is modeled according to the basis of \cite{hughes2010dimension}, using the eigenvectors associated with all of the positive eigenvalues of $\mbf{P}_c \mbf{B} \mbf{P}_c$. The fixed effects parameters $\mbv{\beta}$ are given a $\mbox{MVN}(\mbf{0}, 100^2 I_2)$ prior, and $\tau$ is given an $\hbox{Unif}(0.01,100)$ prior, both of which are intentionally vague. The parametric model uses identical prior specifications.

\begin{figure}
\caption{\baselineskip=10pt The difference of the squared deviations $(Y_i-\widehat{Y}_{(-i)})^2$ for each location of the SHEL model versus the parametric model for the SIDS dataset. The circle indicates Anson county.}
\label{Fi:SIDS}
\begin{center}
\begin{tabular}{cc}
\includegraphics[width=70mm, height=70mm,angle=-90]{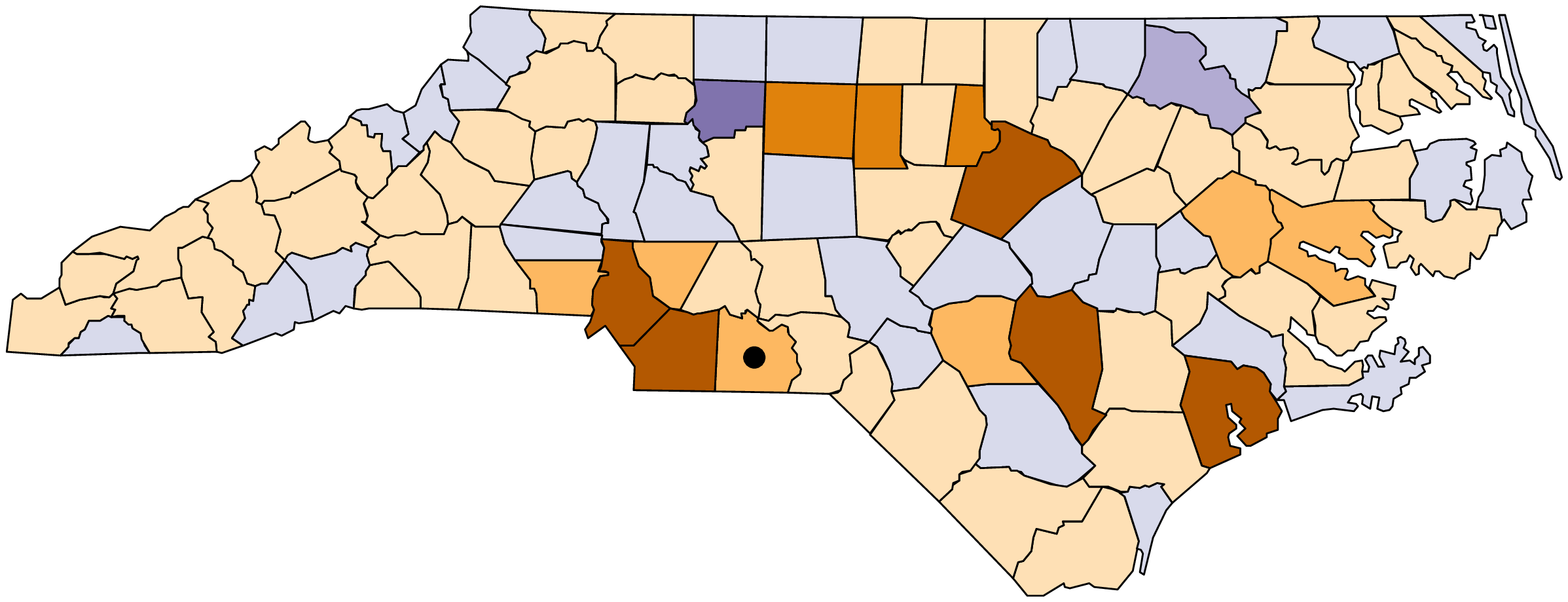} &
\includegraphics[width=70mm, height=70mm,angle=-90]{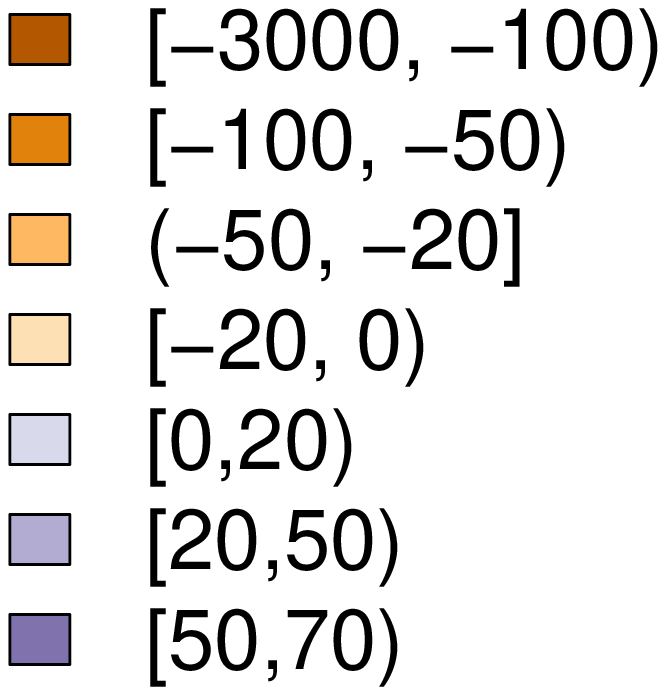} \\
\end{tabular}
\end{center}
\end{figure}

Parameter posterior summaries and the results of a leave-one-out MSPE experiment can be found in Table~\ref{TA:SIDS}. The results show similar medians for the posteriors of the parameters, but the credible interval for $\tau$ (the spatial precision parameter) in the overdispersed Poisson is much larger than the SHEL model. Additionally, the leave-one-out MSPE for the SHEL model is  12.0, approximately 78\% lower than the MSPE of 54.4 for the parametric model. Additionally, not only does the parametric model poorly estimate Anson county, but it also poorly estimates the counties adjacent to it. The SHEL model is clearly more accurate in out-of-sample prediction. This is a case of SHEL methodology fitting the data much better when a parametric model appears to be suggested by the data. The results for both models are based on 10,000 MCMC iterations after 1,000 iterations of burn-in. Convergence was assessed through visual inspection of the sample chains, with no deviations from convergence detected. The results are displayed in Figure~\ref{Fi:SIDS}, and demonstrate that the SHEL model provides superior estimates in terms of MSPE in the majority of locations, but especially in Anson county and the surrounding region.

\subsection{North American Breeding Birds Survey} \label{data:Birds}
Counts of mourning doves in and near Missouri in 2007 from the North American Breeding Bird Survey represent a highly overdispersed spatially point referenced count dataset (mean=30.8, variance=221.7). Counts are collected on 44 sampling routes containing 50 stops each. All routes are 39.2 km in length and each count is assigned to the centroid of the route \citep[see][for a comprehensive description]{BBS}. These data have been previously analyzed using a generalized linear mixed model (GLMM) framework by considering an overdispersed Poisson outcome, with the amount of overdispersion dictated by a latent Gaussian spatial process with the covariance parameterization found in Section~\ref{sec:SS} \citep{gelfand2010handbook}. Modeling is performed equivalently to the simulation study in the previous section in terms of both the model and prior specifications. The results for both models are based on 10,000 MCMC iterations after 1,000 iterations of burn-in, again convergence was assessed through visual inspection of the sample chains with no deviation from convergence detected.

\begin{figure}
\caption{\baselineskip=10pt The difference of the squared deviations $(Y_i-\widehat{Y}_{(-i)})^2$ for each location of the SHEL model and the parametric model for the North American Breeding Birds Survey example.}
\label{Fi:dove}
\begin{center}
\begin{tabular}{cc}
\includegraphics[width=70mm, height=70mm,angle=-90]{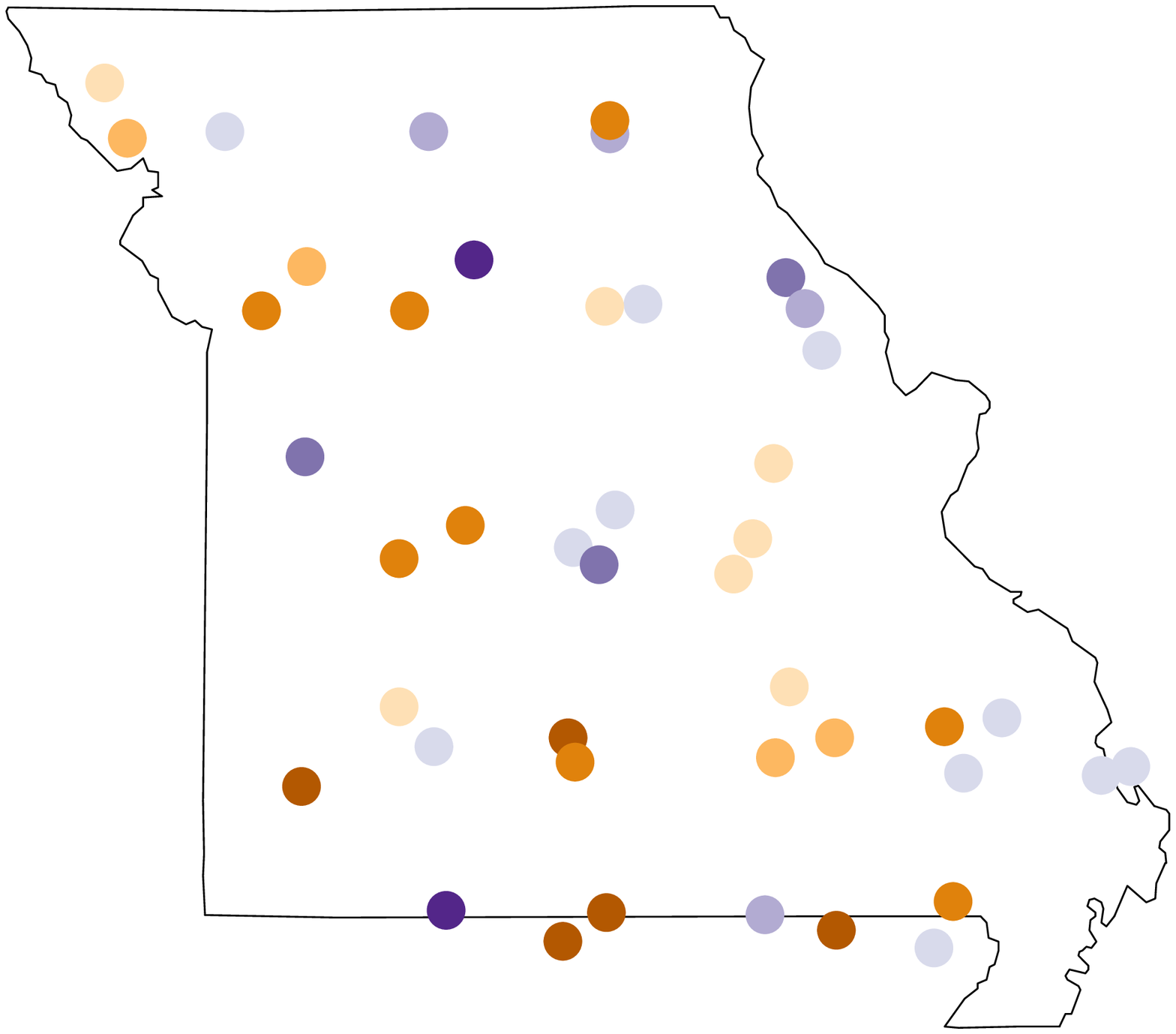} &
\includegraphics[width=70mm, height=70mm,angle=-90]{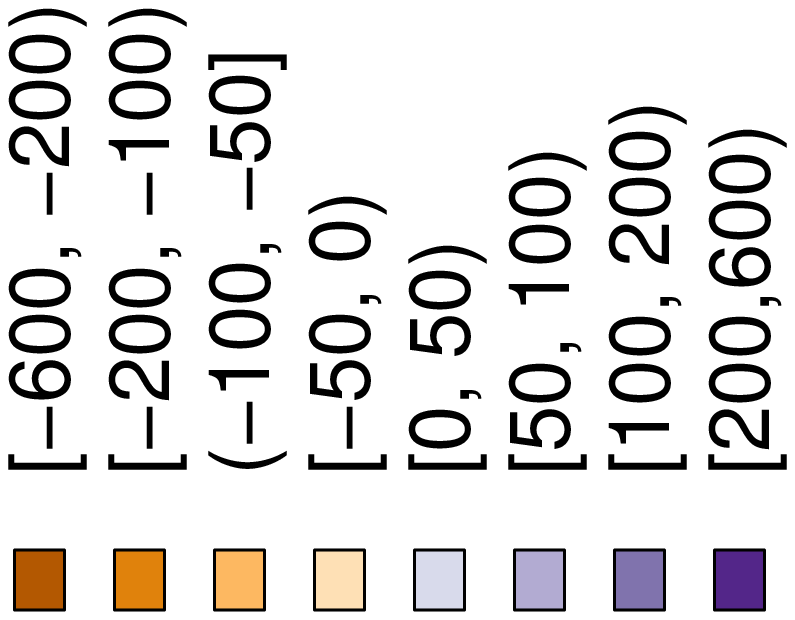}
\end{tabular}
\end{center}
\end{figure}

Results from both the parametric and SHEL models can be found in Table~\ref{TA:Dove}. There are two main differences in the model outcomes. The point estimates of the SHEL model indicate lower spatial variance as well as increasing spatial decay as compared to the parametric model.  This would argue that the SHEL analysis detects less spatial structure than the parametric method. It is worth noting that this is likely due to the flexibility of the empirical data model, which serves to account for some of spatial structure of the data. Secondly, we again see an improved predictive ability of the SHEL framework for these data, as indicated by the leave-one-out MSPEs. This decrease is noticeable, with the leave-one-out MSPE for the SHEL model being 195.4, nearly a 15\% reduction over the 228.5 for parametric model. The results are displayed in Figure~\ref{Fi:dove} and again demonstrate superior prediction in terms of MSPE in the majority of locations analyzed.

\section{Discussion}\label{sec:Disc}
In this paper, we have proposed a general framework for including empirical data models in the BHM framework. We have shown that the SHEL model can explicitly accommodate spatial correlation on irregular lattices as well as handle spatial point-referenced data not collected on a regular grid, both of which are novel models. The simulation study in Section~\ref{sec:SS} demonstrated improved predictive performance and corroborated the results for the spatial point-referenced North American Breeding Birds Survey data presented in Section~\ref{sec:IE}.  In order for the models we propose to be useful in practice, we have provided detailed discussion regarding sampling and computational considerations.

Importantly, we have shown that the SHEL framework outperforms standard parametric analyses in three distinct and unrelated case studies. In every case, the SHEL model has outperformed parametric models in terms of out of sample prediction as measured by reduction in MSPE of at least 15\%. In the case of the SIDS data and the ACS data, we have outperformed a standard analyses by a reduction of 30\% in terms of MSPE.  While the SHEL paradigm can certainly be used for inference, EL methods are known to produce asymptotic credible intervals that slightly undercover the true parameter values in the mean structure of multiple regression models \citep{fang2006empirical}. Therefore, one should take care when interpreting the credible intervals produced by such methods.

The SHEL model overcomes one of the main difficulties in standard EL analysis, which is handling dependence in the outcomes. That is, the SHEL model places the dependence structure at the process and parameter stages of the hierarchy. This makes the framework extremely advantageous for a wide range of problems where parametric modeling assumptions may be difficult to verify.  Accordingly, the SHEL model  provides a unified BHM framework that is capable of handling a broad range of dependence structures, including spatial dependence, as illustrated here.  In short, by casting the SHEL model within the BHM paradigm we provide an extremely flexible approach that takes advantage of conditional thinking and is, therefore, capable of effectively modeling parameters.  In addition, as a byproduct of the BHM specification, we are easily able incorporate relevant scientific information, while providing a quantification of uncertainty of our predictions.

\section*{Acknowledgments}
This research was partially supported by the U.S. National Science Foundation (NSF) and the U.S. Census Bureau under NSF grant SES-1132031, funded through the NSF-Census Research Network (NCRN) program.

\section*{Appendix A: Proof of Theorem 1}\label{AppendixProof}
\begin{appendix}
\Appendix    
\renewcommand{\theequation}{A.\arabic{equation}}
\setcounter{equation}{0}

Let $\mathcal{C}(\mbf{A})$ denote the column space of a matrix A and $\mathcal{N}(\mbf{A})$ represent the null space. Assume that $\mbf{X}$ contain a column equal to the one vector, which implies that the model contains an intercept. We proceed by contradiction. First, suppose there exists $\mbf{v} \neq \mbf{0}$ such that
\begin{equation}
\nonumber \mbf{v}'\mbf{M}'(\mbf{B}_+-\mbf{B})\mbf{Mv}=0.
\end{equation}
Let $\mbf{P} \mbv{\Lambda} \mbf{P}'$ represent the eigenspace decomposition of $(\mbf{B}_+-\mbf{B})$.
Then we have
\begin{equation}
\nonumber \mbf{v}'\mbf{M}'\mbf{P} \mbv{\Lambda}^{\frac{1}{2}} \mbv{\Lambda}^{\frac{1}{2}} \mbf{P}'\mbf{Mv}=0,
\end{equation}
for some $\mbf{v} \neq \mbf{0}$. This implies
\begin{equation}
\mbv{\Lambda}^{\frac{1}{2}} \mbf{P}'\mbf{Mv}=0
\label{eqn:posdefspec}
\end{equation}
for this choice of $\mbf{v}$.
Now, we know that, for the ICAR specification we have chosen,
\begin{eqnarray}
\nonumber 1 &=&\hbox{nullity}(\mbf{B}_+-\mbf{B})\\
\nonumber &=&\hbox{nullity}(\mbf{P} \mbv{\Lambda}^{\frac{1}{2}} \mbv{\Lambda}^{\frac{1}{2}} \mbf{P}') \\
&\geq& \hbox{nullity}(\mbv{\Lambda}^{\frac{1}{2}} \mbf{P}').
\label{eqn:nullsize}
\end{eqnarray}
Note that
\begin{eqnarray}
\nonumber & &\mathbf{1}'(\mbf{B}_+-\mbf{B}) \mathbf{1}=0 \\
\nonumber  & &\Rightarrow \mathbf{1}'(\mbf{P} \mbv{\Lambda}^{\frac{1}{2}} \mbv{\Lambda}^{\frac{1}{2}} \mbf{P}')\mathbf{1}=0 \\
& &\Rightarrow \mbv{\Lambda}^{\frac{1}{2}} \mbf{P}'\mathbf{1}=0.
\label{eqn:nullcontains}
\end{eqnarray}
Together with (\ref{eqn:nullsize}), (\ref{eqn:nullcontains}) implies $\mathcal{N}(\mbv{\Lambda}^{\frac{1}{2}}\mbf{P}')=\{\mbf{0},\mbf{1}\}$, as nullity$(\mbv{\Lambda}^{\frac{1}{2}}\mbf{P}') \leq 1$ and we have demonstrated that $\mbv{\Lambda}^{\frac{1}{2}} \mbf{P}'\mathbf{1}=\mbf{0}$.
$\mbf{M}$ is full rank by construction; so, $\mbf{Mv}\neq \mbf{0}$ for $\mbf{v}\neq\mbf{0}$. So, if $\mbf{v}'\mbf{M}'(\mbf{B}_+-\mbf{B})\mbf{Mv}=0$, which in turn implies $\mbv{\Lambda}^{\frac{1}{2}} \mbf{P}'\mbf{Mv}=0$, we must have that $\mbf{1} \in \mathcal{C}(\mbf{M})$ if $\mbf{v} \neq 0$.
However, $\mathcal{C}(\mbf{M}) \perp \mathcal{C}(\mbf{X})$ and $\mbf{1} \in \mathcal{C}(\mbf{X})$, which is a contradiction. Therefore, $\mbf{v}'\mbf{M}'(\mbf{B}_+-\mbf{B})\mbf{Mv}=0$ implies $\mbf{v}=\mbf{0}$, and we have that $\mbf{M}'(\mbf{B}_+-\mbf{B})\mbf{M}$ is positive definite.

\section*{Appendix B: MCMC Sampling Algorithm}\label{AppendixSampling}

\renewcommand{\theequation}{B.\arabic{equation}}
\setcounter{equation}{0}

Herein, we provide the sampling algorithm used to sample the SHEL Fay-Herriot model. Sampling algorithms for the other models discussed are similar and proceed in a straightforward manner. The sampling algorithm proceeds as follows.
\begin{enumerate}
\item Utilizing the estimating equations
\begin{eqnarray}
\nonumber & & \sum_{i=1}^{n} w_{i}\{z_{i}-\mbf{x}_i' \mbv{\beta} \}=0, \\
\nonumber & & \sum_{i=1}^{n} \{w_{i}(z_{i}-\mbf{x}_i' \mbv{\beta})^2/\sigma_i^2\}-1 =0,
\end{eqnarray}
and the gmm package in the R programming language, generate the MELEs for $\mbv{\beta}$ given that the latent process, $\mbf{Y}_q^*$, is set identically equal to zero. Next set the initial values for $\mbv{\beta}$ to the MELE values and set $\mbf{Y}_q^*=\mbf{0}$. This provides starting values for $\mbv{\theta}=\mbf{X}\mbv{\beta}+\mbf{M}\mbf{Y}_q^*$ that generate a set of weights $\{w_i\}$ guaranteed to be in the simplex
\begin{equation}
  \label{eq:EEsimplexapp}
 W_{\theta}=\left\{ \sum_{i=1}^{n} w_{i} =1; w_{i}>0 \text{ for all $i$ }; \sum_{i=1}^{n}w_{i}m_{j}(z_i,\mbv{\theta})=0 \hbox{ for all $j$} \right\}.
\end{equation}

\item \textbf{Sampling } $\mbf{Y}_q^*$

In blocks of size $B$ (we use $B$=15) we sample $\mbf{Y}_q^*$ using a random walk Metropolis-Hastings step with a multivariate normal for block $k$, $\widetilde{\mbf{y}}_{q,k}^* \sim N(\mbf{y}_{q,k}^*,\Sigma_{Y_{q,k}})$, where the proposal covariance $\Sigma_{Y_{q,k}}$ is tuned based on pilot chains \citep{gelman2013bayesian}. We utilize the proposed values with the estimating equations
\begin{eqnarray}
\nonumber & &\sum_{i=1}^{n} w_{i}\{z_{i}-\mbf{x}_i' \mbv{\beta} -\mbf{M}_i\widetilde{\mbf{y}_q}^*\}=0 \\
\nonumber & &\sum_{i=1}^{n} \{w_{i}(z_{i}-\mbf{x}_i' \mbv{\beta} -\mbf{M}_i\widetilde{\mbf{y}_q}^*)^2/\sigma_i^2\}-1 =0
\end{eqnarray}
to generate a set of weights $\{\widetilde{w}_i\}$, where $\mbf{M}_i$ is the $i$-th row of $\mbf{M}$, and the elements of $\widetilde{\mbf{y}_q}^*$ in block $k$ are set to $\{\widetilde{\mbf{y}}_{k,q}^*\}$, and the elements of $\widetilde{\mbf{y}_q}^*$ that are not in block $k$ are left as $\{\mbf{y}_q^*\}$. Once generated, we verify that $\{\widetilde{w}_i\}$ satisfies (\ref{eq:EEsimplexapp}). If it does not, the block of $B$ elements of $\{\mbf{y}_q^*\}$ remains at their previous values, and we move to the next block of $B$ elements of $\{\mbf{y}_q^*\}$. Otherwise, perform a Metropolis-Hastings step with the posterior density ratio
\begin{eqnarray}
\nonumber & & \Upsilon_{\mbf{y}_q}=\frac{p(\mbf{Z}|\widetilde{\mbf{y}_q}^*,\mbf{\beta})\pi(\widetilde{\mbf{y}_q}^*|\tau)}{p(\mbf{Z}|\mbf{y}_q^*,\mbf{\beta})\pi(\mbf{y}_q^*|\tau)} \\
\nonumber & & \Upsilon_{\mbf{y}_q}=\frac{\prod_{i=1}^n (\widetilde{w}_i) \exp(-\frac{1}{2}\widetilde{\mbf{y}_q}^{*'}\mbf{M}' \{\mbf{B}_+ -\mbf{B}\} M \widetilde{\mbf{y}_q}^* \tau)}{ \prod_{i=1}^n (w_i) \exp(-\frac{1}{2}\mbf{y}_q^{*'} \mbf{M}' \{\mbf{B}_+ -\mbf{B}\} M \mbf{y^*} \tau)}
\end{eqnarray}
We accept $\widetilde{\mbf{y}_q}^*$ if $\Upsilon_{\mbf{y}_q}>u_{\mbf{y}_q}$, where $u_{\mbf{y}_q} \sim \hbox{Unif}(0,1)$.
Repeat this process for every block of $B$ elements of $\{y^*_q\}$ until the entire set has been considered.

\item \textbf{Sampling } $\mbv{\beta}$

We sample $\mbv{\beta}$ using a random walk Metropolis-Hastings step with a multivariate normal proposal $\widetilde{\mbv{\beta}} \sim N(\mbv{\beta},\Sigma_{\beta})$, where the proposal covariance $\Sigma_\beta$ is tuned based on pilot chains. We use the estimating equations
\begin{eqnarray}
\nonumber & &\sum_{i=1}^{n} w_{i}\{z_{i}-\mbf{x}_i' \widetilde{\mbv{\beta}} -\mbf{M}_i \mbf{y}_q^* \}=0 \\
\nonumber & &\sum_{i=1}^{n} \{w_{i}(z_{i}-\mbf{x}_i' \widetilde{\mbv{\beta}} -\mbf{M}_i \mbf{y}_q^*)^2/\sigma_i^2\}-1 =0
\end{eqnarray}
to generate a set of weights $\{\widetilde{w}_i\}$. Once generated, we verify that $\{\widetilde{w}_i\}$ satisfies (\ref{eq:EEsimplexapp}). If it does not, we set $\mbv{\beta}$ to the previous values. Otherwise, perform a Metropolis-Hastings step with the posterior density ratio
\begin{eqnarray}
\nonumber & &\Upsilon_{\beta}=\frac{p(\mbf{Z}|{\mbf{y}_q^*},\widetilde{\mbf{\beta}})\pi(\widetilde{\mbv{\beta}}|\tau)}{p(\mbf{Z}|\mbf{y}_q^*,\mbf{\beta})\pi(\mbv{\beta}|\tau)} \\
\nonumber & &\Upsilon_{\beta}=\frac{\prod_{i=1}^n (\widetilde{w}_i) \exp(-\frac{1}{2}\{\widetilde{\mbv{\beta}}-\mbv{\beta}^{*'}\} \{\widetilde{\mbv{\beta}}-\mbv{\beta}_{\hbox{\tiny{WLS}}}\} g \tau)}{\prod_{i=1}^n (w_i) \exp( -\frac{1}{2}\{\mbv{\beta}-\mbv{\beta}^{*'}\} \{\mbv{\beta}-\mbv{\beta}_{\hbox{\tiny{WLS}}}\} g \tau)}
\end{eqnarray}
where $g$ is Zellner's g prior and $\mbv{\beta}_{\hbox{\tiny{WLS}}}$ are the weighted least squares estimates of $\mbv{\beta}$.
We accept $\widetilde{\mbv{\beta}}$ if $\Upsilon_{\beta}>u_\beta$, where $u_\beta \sim \hbox{Unif}(0,1)$.

\item \textbf{Sampling } $\tau$

We sample $\tau$ using a random walk Metropolis-Hastings step with a normal proposal $\widetilde{\tau}^* \sim N(\tau,\Sigma_{\tau})$, where the proposal variance $\Sigma_\tau$ is tuned based on pilot chains. We then perform a Metropolis-Hastings with the posterior density ratio
\begin{eqnarray}
\nonumber & & \Upsilon_{\tau}=\frac{\pi(\mbv{\beta}|\widetilde{\tau})\pi(\mbf{Y}_q^*|\widetilde{\tau})\pi(\widetilde{\tau})} {\pi(\mbv{\beta}|\tau)\pi(\mbf{Y}_q^*|\tau)\pi(\tau)}\\
\nonumber  & &\Upsilon_\tau=\frac{\widetilde{\tau}^{\frac{q+p}{2}}\exp(-\frac{1}{2}\mbf{y}_q^{*'} \mbf{M}' \{\mbf{B}_+ -\mbf{B}\} M \mbf{y}_q^*\widetilde{\tau}) \exp(-\frac{1}{2}\{\mbv{\beta}-\mbv{\beta}^{*'}\} \{ \mbv{\beta}-\mbv{\beta}_{\hbox{\tiny{WLS}}}\} \widetilde{\tau} g ) \widetilde{\tau}^{- (1+\alpha_1)} \exp(-\frac{\alpha_2}{\widetilde{\tau}})}{
\tau^{\frac{q+p}{2}}\exp(-\frac{1}{2}\mbf{y}_q^{*'} \mbf{M}' \{\mbf{B}_+ -\mbf{B}\} M \mbf{y}_q^* \tau) \exp(-\frac{1}{2}\{\mbv{\beta}-\mbv{\beta}^{*'}\} \{ \mbv{\beta}-\mbv{\beta}_{\hbox{\tiny{WLS}}}\} \tau g ) \tau^{- (1+\alpha_1)} \exp(-\frac{\alpha_2}{\tau})}
 \end{eqnarray}
where we have used an IG($\alpha_1,\alpha_2$) prior distribution for $\tau$.
We accept $\widetilde{\tau}$ if $\Upsilon_{\tau}>u_{\tau}$, where $u_{\tau} \sim \hbox{Unif}(0,1)$.

\item Utilizing (\ref{eq:EEsimplexapp}), steps 2--4 are repeated until convergence.

\end{enumerate}
\end{appendix}
\bibliographystyle{jasa}
\bibliography{STSN}

\begin{thebibliography}{40}
\newcommand{\enquote}[1]{``#1''}
\expandafter\ifx\csname natexlab\endcsname\relax\def\natexlab#1{#1}\fi

\bibitem[\protect\citename{Bandyopadhyay et~al.,
  }2012]{bandyopadhyay2012frequency}
Bandyopadhyay, S., Lahiri, S.~N., and Nordman, D. (2012).
\newblock \enquote{Frequency domain empirical likelihood method for irregularly
  spaced spatial data.}
\newblock {\em Unpublished manuscript, Lehigh University, PA, USA\/}.
\newblock Http://www.lehigh.edu/~sob210/SFDEL-AOS2.pdf.

\bibitem[\protect\citename{Berliner, }1996]{berliner1996hierarchical}
Berliner, L.~M. (1996).
\newblock \enquote{Hierarchical {B}ayesian time series models.}
\newblock In {\em Maximum entropy and {B}ayesian methods\/},  15--22. Springer.

\bibitem[\protect\citename{Besag et~al., }1991]{besag1991}
Besag, J., York, J., and Molli\'{e}, A. (1991).
\newblock \enquote{{B}ayesian image restoration with two applications in
  spatial statistics (with discussion).}
\newblock {\em Annals of the Institute of Statistical Mathematics\/}, 43,
  1--59.

\bibitem[\protect\citename{Chang and Mukerjee, }2008]{chang2008bayesian}
Chang, I. and Mukerjee, R. (2008).
\newblock \enquote{{B}ayesian and frequentist confidence intervals arising from
  empirical-type likelihoods.}
\newblock {\em Biometrika\/}, 95, 1, 139--147.

\bibitem[\protect\citename{Chaudhuri and Ghosh, }2011]{chaudhuri2011empirical}
Chaudhuri, S. and Ghosh, M. (2011).
\newblock \enquote{Empirical likelihood for small area estimation.}
\newblock {\em Biometrika\/}, 98, 2, 473--480.

\bibitem[\protect\citename{Chauss{\'e}, }2010]{chausse2010computing}
Chauss{\'e}, P. (2010).
\newblock \enquote{Computing Generalized Method of Moments and Generalized
  Empirical Likelihood with R.}
\newblock {\em Journal of Statistical Software\/}, 34, 11, 1--35.

\bibitem[\protect\citename{Chen et~al., }2002]{chen2002using}
Chen, J., Sitter, R., and Wu, C. (2002).
\newblock \enquote{Using empirical likelihood methods to obtain range
  restricted weights in regression estimators for surveys.}
\newblock {\em Biometrika\/}, 89, 1, 230--237.

\bibitem[\protect\citename{Cressie and Chan, }1989]{cressie1989spatial}
Cressie, N. and Chan, N.~H. (1989).
\newblock \enquote{Spatial modeling of regional variables.}
\newblock {\em Journal of the American Statistical Association\/}, 84, 406,
  393--401.

\bibitem[\protect\citename{Cressie and Wikle, }2011]{wiklecressie}
Cressie, N. and Wikle, C.~K. (2011).
\newblock {\em Statistics for Spatio-Temporal Data\/}.
\newblock Hoboken, NJ: John Wiley and Sons.

\bibitem[\protect\citename{Fang and Mukerjee, }2006]{fang2006empirical}
Fang, K. and Mukerjee, R. (2006).
\newblock \enquote{Empirical-type likelihoods allowing posterior credible sets
  with frequentist validity: Higher-order asymptotics.}
\newblock {\em Biometrika\/}, 93, 3, 723--733.

\bibitem[\protect\citename{Fay and Herriot, }1979]{fay1979estimates}
Fay, R. and Herriot, R. (1979).
\newblock \enquote{Estimates of income for small places: an application of
  {J}ames-{S}tein procedures to census data.}
\newblock {\em Journal of the American Statistical Association\/}, 74,
  269--277.

\bibitem[\protect\citename{Gelman et~al., }2013]{gelman2013bayesian}
Gelman, A., Carlin, J.~B., Stern, H.~S., Dunson, D.~B., Vehtari, A., and Rubin,
  D.~B. (2013).
\newblock {\em Bayesian {D}ata {A}nalysis\/}.
\newblock 3rd ed. CRC Press, Boca Raton, FL.

\bibitem[\protect\citename{Getis and Ord, }1992]{getis1992}
Getis, A. and Ord, J. (1992).
\newblock \enquote{The Analysis of Spatial Association by Use of Distance
  Statistics.}
\newblock {\em Geographical Analysis\/}, 23, 3, 190--205.

\bibitem[\protect\citename{Hughes and Haran, }2013]{hughes2010dimension}
Hughes, J. and Haran, M. (2013).
\newblock \enquote{Dimension reduction and alleviation of confounding for
  spatial generalized linear mixed models.}
\newblock {\em Journal of the Royal Statistical Society: Series B (Statistical
  Methodology)\/}, 75, 1, 139--159.

\bibitem[\protect\citename{Kaiser and Nordman, }2012]{kaiser2012blockwise}
Kaiser, M.~S. and Nordman, D.~J. (2012).
\newblock \enquote{Blockwise empirical likelihood for spatial Markov model
  assessment.}
\newblock {\em Unpublished manuscript\/}.
\newblock Http://streaming.stat.iastate.edu/~stat506/papers/SBEL.pdf.

\bibitem[\protect\citename{Kitamura, }1997]{kitamura1997empirical}
Kitamura, Y. (1997).
\newblock \enquote{Empirical likelihood methods with weakly dependent
  processes.}
\newblock {\em The Annals of Statistics\/}, 25, 5, 2084--2102.

\bibitem[\protect\citename{Kolaczyk, }1994]{kolaczyk1994}
Kolaczyk, E.~D. (1994).
\newblock \enquote{Empirical Likelihood for Generalized Linear Models.}
\newblock {\em Statistica Sinica\/}, 4, 199--218.

\bibitem[\protect\citename{Kulldorf, }1997]{kulldorf1997}
Kulldorf, M. (1997).
\newblock \enquote{A spatial scan statistic.}
\newblock {\em Communications in Statistics - Theory and Methods\/}, 26, 6,
  1481--1496.

\bibitem[\protect\citename{Lazar, }2003]{lazar2003bayesian}
Lazar, N. (2003).
\newblock \enquote{{B}ayesian empirical likelihood.}
\newblock {\em Biometrika\/}, 90, 2, 319--326.

\bibitem[\protect\citename{Monahan and Boos, }1992]{monahan1992proper}
Monahan, J. and Boos, D. (1992).
\newblock \enquote{Proper likelihoods for {B}ayesian analysis.}
\newblock {\em Biometrika\/}, 79, 2, 271--278.

\bibitem[\protect\citename{Newey and Smith, }2004]{newey2004higher}
Newey, W. and Smith, R. (2004).
\newblock \enquote{Higher order properties of GMM and generalized empirical
  likelihood estimators.}
\newblock {\em Econometrica\/}, 72, 1, 219--255.

\bibitem[\protect\citename{Nordman, }2008]{nordman2008empirical}
Nordman, D. (2008).
\newblock \enquote{An empirical likelihood method for spatial regression.}
\newblock {\em Metrika\/}, 68, 3, 351--363.

\bibitem[\protect\citename{Nordman and Caragea, }2008]{nordman2008point}
Nordman, D.~J. and Caragea, P.~C. (2008).
\newblock \enquote{Point and interval estimation of variogram models using
  spatial empirical likelihood.}
\newblock {\em Journal of the American Statistical Association\/}, 103, 481,
  350--361.

\bibitem[\protect\citename{Owen, }1988]{owen1988empirical}
Owen, A. (1988).
\newblock \enquote{Empirical likelihood ratio confidence intervals for a single
  functional.}
\newblock {\em Biometrika\/}, 75, 2, 237--249.

\bibitem[\protect\citename{Owen, }2001]{owen2001}
Owen, A.~B. (2001).
\newblock {\em Empirical Likelihood\/}.
\newblock Chapman and Hall / CRC.
\newblock Boca Raton, FL.

\bibitem[\protect\citename{Qin and Lawless, }1994]{qin1994empirical}
Qin, J. and Lawless, J. (1994).
\newblock \enquote{EMPIRICAL LIKELIHOOD AND GENERAL ESTIMATING EQUATIONS.}
\newblock {\em The Annals of Statistics\/}, 22, 1, 300--325.

\bibitem[\protect\citename{{R Core Team}, }2013]{Rlanguage}
{R Core Team} (2013).
\newblock {\em R: A Language and Environment for Statistical Computing\/}.
\newblock R Foundation for Statistical Computing, Vienna, Austria.

\bibitem[\protect\citename{Reich et~al., }2006]{reich2006effects}
Reich, B., Hodges, J., and Zadnik, V. (2006).
\newblock \enquote{Effects of Residual Smoothing on the Posterior of the Fixed
  Effects in Disease-Mapping Models.}
\newblock {\em Biometrics\/}, 62, 4, 1197--1206.

\bibitem[\protect\citename{Robbins et~al., }1986]{BBS}
Robbins, C., Bystrak, D., and Geissler, P. (1986).
\newblock \enquote{The Breeding Birds Survey: Its First Fifteen Years,
  1965-1979.}
\newblock {\em {USDOI}, Fish and Wildlife Resource Publication 157\/}.
\newblock Washington, D.C.

\bibitem[\protect\citename{Rue and Held, }2005]{rue2005gaussian}
Rue, H. and Held, L. (2005).
\newblock {\em Gaussian {M}arkov {R}andom {F}ields: {T}heory and
  {A}pplications\/}.
\newblock Boca Raton, FL: Chapman \& Hall/CRC.

\bibitem[\protect\citename{Schennach, }2005]{schennach2005bayesian}
Schennach, S. (2005).
\newblock \enquote{{B}ayesian exponentially tilted empirical likelihood.}
\newblock {\em Biometrika\/}, 92, 1, 31--46.

\bibitem[\protect\citename{Schennach, }2007]{schennach2007point}
--- (2007).
\newblock \enquote{Point estimation with exponentially tilted empirical
  likelihood.}
\newblock {\em The Annals of Statistics\/}, 35, 2, 634--672.

\bibitem[\protect\citename{Sengupta and Cressie, }2013]{sengupta2013}
Sengupta, A. and Cressie, N. (2013).
\newblock \enquote{Empirical Hierarchical Modelling for Count Data using the
  Spatial Random Effects Model.}
\newblock {\em Spatial Economic Analysis\/}, 8, 3, 389--418.

\bibitem[\protect\citename{Smith, }1997]{smith1997alternative}
Smith, R. (1997).
\newblock \enquote{Alternative Semi-parametric Likelihood Approaches to
  Generalised Method of Moments Estimation.}
\newblock {\em The Economic Journal\/}, 107, 441, 503--519.

\bibitem[\protect\citename{Symons et~al., }1983]{symons1983}
Symons, M.~J., Grimson, R.~C., and Yuan, Y.~C. (1983).
\newblock \enquote{Clustering of Rare Events.}
\newblock {\em Biometrics\/}, 39, 1, 193--205.

\bibitem[\protect\citename{Wall, }2004]{wall2004close}
Wall, M. (2004).
\newblock \enquote{A close look at the spatial structure implied by the {CAR}
  and {SAR} models.}
\newblock {\em Journal of Statistical Planning and Inference\/}, 121, 2,
  311--324.

\bibitem[\protect\citename{Wikle, }2010]{gelfand2010handbook}
Wikle, C. (2010).
\newblock \enquote{Hierarchical Modeling with Spatial Data.}
\newblock In {\em Handbook of Spatial Statistics\/}, eds. A.~Gelfand, P.~J.
  Diggle, P.~Guttorp, and M.~Fuentes. CRC Press.
\newblock Boca Raton, FL.

\bibitem[\protect\citename{Wikle and Berliner, }2005]{wikleberliner2005}
Wikle, C. and Berliner, L. (2005).
\newblock \enquote{Combining {I}nformation {A}cross {S}patial {S}cales.}
\newblock {\em Techonmetrics\/}, 47, 80--91.

\bibitem[\protect\citename{Wikle, }2003]{wikle2003hierarchical}
Wikle, C.~K. (2003).
\newblock \enquote{Hierarchical {B}ayesian models for predicting the spread of
  ecological processes.}
\newblock {\em Ecology\/}, 84, 6, 1382--1394.

\bibitem[\protect\citename{Zellner, }1986]{zellner1986bayesian}
Zellner, A. (1986).
\newblock \enquote{{B}ayesian estimation and prediction using asymmetric loss
  functions.}
\newblock {\em Journal of the American Statistical Association\/}, 81, 394,
  446--451.

\end{thebibliography}

\pagebreak

\begin{table}\footnotesize
\begin{center}
\begin{tabular}{|c|c|c|c|c|}
\hline
Model & $\beta_{0}$ & $\beta_1$ & $A$ & MSPE \\ \hline
SHEL &2.164 & -0.042 &0.287  & 0.066\\
 & (2.051, 2.256) & (-0.063, -0.015) & (0.157, 0.628)& \\ \hline
Independence EL & 2.230 & -0.077 & 0.008 & 0.128 \\
& (2.210, 2.364) & (-0.095, -0.058) & (0.004, 0.015) & \\ \hline
DP EL & 2.331 & -0.0375 & 0.049 & 0.128 \\
& (2.170, 2.474) & (-0.069, -0.002) & (0.006, 0.745) & \\ \hline
Independence Parametric & 2.094 & -0.006 & 0.142 & 0.130 \\
& (1.971, 2.217) & (-0.027, 0.015) & (0.109 0.187) & \\ \hline
Spatial Parametric & 2.327 & -0.058 & 0.503 & 0.076 \\
& (2.284, 2.370) & (-0.067, -0.050) & (0.345, 0.765) & \\ \hline
\end{tabular}
\parbox{5in}{
\caption{\baselineskip=10pt Posterior medians and 95\% (central) credible intervals for the FH example (Section 5.1). $A$ represents the variance of $\mbf{y}$ in the Chaudhuri and Ghosh (2011) parameterizations, and $\tau^{-1}$ for the SHEL parameterization. MPV is the mean posterior variance of $\mbv{\theta}$ for each model.}
\label{TA:FH}
}
\end{center}
\end{table}

\begin{table}\footnotesize
\begin{center}
\begin{tabular}{|c|c|c|c|c|}
\hline
Model & $\beta_{0}$ & $\beta_{1}$ & $\tau$ &MSPE \\ \hline
Parametric  &-1.071 &1.899 & 1.102 &54.4 \\
 & (-1.441, -0.724) & (1.322,2.494) & (0.602, 2.050) &\\ \hline
SHEL  & -0.971 & 1.723& 0.289& 12.0 \\
 & (-1.540, -0.404) & (0.794, 2.659) & (0.142, 0.635)  & \\ \hline
\end{tabular}
\parbox{5in}{
\caption{\baselineskip=10pt Posterior medians and 95\% (central) credible intervals for the SIDS example (Section 5.2).}
\label{TA:SIDS}
}
\end{center}
\end{table}

\begin{table}\footnotesize
\begin{center}
\begin{tabular}{|c|c|c|c|c|}
\hline
Model & $\beta_{0}$ & $\sigma_u^2$ &$\phi$ &MSPE \\ \hline
Parametric  &3.322 & 0.377 & 0.587 & 228.5\\
 & (3.261 3.384) & (0.157, 1.272) & (0.028, 2.937)& \\ \hline
SHEL  & 3.390& 0.230 & 1.580&  195.4 \\
 & (3.277, 3.503) & (0.042, 0.751) & (0.082, 3.868)  & \\ \hline
\end{tabular}
\parbox{5in}{
\caption{\baselineskip=10pt Posterior medians and 95\% (central) credible intervals for the North American Breeding Bird Survey example (Section 5.3).}
\label{TA:Dove}
}
\end{center}
\end{table}

\baselineskip=14pt \vskip 4mm\noindent

\baselineskip=14pt \vskip 4mm\noindent

\baselineskip=14pt \vskip 4mm\noindent

\end{document}